\documentclass[a4paper,11pt]{article}
\usepackage{pos}

\title{Chiral Lagrangian for Karsten-Wilczek Minimally Doubled Fermions}

\author*[a]{Kunal Shukre}
\author[b]{Dipankar Chakrabarti}
\author[a]{Subhasish Basak}

\affiliation[a]{School of Physical Sciences, National Institute of Science Education and Research\\
Bhubaneswar and Homi Bhabha National Institute, India}

\affiliation[b]{Department of Physics, Indian Institute of Technology Kanpur, India}

\emailAdd{kunalsatyajit.shukre@niser.ac.in}
\emailAdd{dipankar@iitk.ac.in}
\emailAdd{sbasak@niser.ac.in}

\abstract{Lattice chiral perturbation theory is developed for Karsten-Wilczek fermions, a variant of minimally doubled fermions. As a first step, we consider the näive fermionic field on lattice without its doubler. Once the symmetries of the action, the Symanzik effective theory and the spurion structure are established for the single fermion, we extend our study to include the doubler. Symanzik effective actions are considered up to five-dimensional operators in both cases. The two fermionic tastes are realized by point-splitting the quark wavefunction in the coordinate space. Spurion analysis is used to construct the chiral lagrangians for Karsten-Wilczek fermions from the Symanzik actions. In this work, we have not included a pion that is massive in the continuum limit.}

\FullConference{%
  The XLI International Symposium on Lattice Field Theory\\
  July 28-August 3, 2024\\
  University of Liverpool, Liverpool, United Kingdom
}

\begin{document}
\maketitle

\section{Introduction}

Chiral Perturbation Theory ($\chi$PT) was developed in the early eighties by Gasser and Lleutwyler \cite{gasser1984chiral}. Since then $\chi$PT has been extensively used to explore continuum QCD at low energies and has also become an essential component in extracting physical quantities obtained using unphysical parameters (both lattice spacing and quark mass) from lattice simulations. It describes the low-energy interactions of the pseudoscalar mesons arising as Goldstone modes from the spontaneous breaking of the chiral symmetry of the QCD vacuum. $\chi$PT has also been extended to include the lattice discretization errors. This allows a continuous extrapolation to $a\rightarrow 0$ and $m \rightarrow m_{\text{physical}}$ simultaneously.

Symanzik effective theory \cite{symanzik1983continuum,symanzik1983continuum2} describes lattice actions near the continuum limit using higher dimensional operators in orders of lattice spacing $a$ which appears explicitly in the action. Such a chiral Lagrangian is an expansion in the quark masses (like continuum $\chi$PT) \textit{and} in the lattice spacing. This formalism is known as lattice $\chi$PT. Well-known examples of lattice $\chi$PT are Wilson $\chi$PT \cite{bar2004chiral,rupak2002chiral,sharpe1998spontaneous} and Staggered $\chi$PT \cite{lee1999partial}. In this work we address minimally doubled fermion $\chi$PT, especially for Karsten-Wilczek formalism \cite{karsten1981lattice,wilczek1987lattice}.

According to the Nielsen-Ninomiya No-Go Theorem \cite{nielsen1981no}, fermion actions that have an exact chiral symmetry must have at least two doublers. The actions that satisfy the minimum criteria are known as minimally-doubled fermions. All minimally doubled fermions preserve exact chiral symmetry and locality \cite{creutz2010classification}. One such action with the two doublers on the temporal axis was suggested by Karsten and Wilczek. Another popular variant of minimally doubled fermions is Borici-Creutz fermions \cite{boricci2008minimally,borici2008creutz,creutz2008four,creutz2008local}, whose $\chi$PT construction will be taken up in the future.

The aim of this work is to construct a lattice $\chi$PT for Karsten-Wilczek minimally doubled fermions. We discuss two cases, first is the simplistic one where we do not consider the existence of the two doublers. The second case considers the two doublers or "quark tastes" and splits the quark wavefunction into two. Symanzik effective theory is constructed in both the cases and then the standard process of spurion analysis is used to construct the chiral Lagrangians.
\section{Karsten-Wilczek Fermions}
\label{sec:2}

The Karsten-Wilczek (KW) action \cite{karsten1981lattice,wilczek1987lattice} in presence of $SU(3)$ gauge fields on a discrete 4-dimensional spacetime lattice is written as,
\begin{equation}\label{1}
\begin{aligned}    S_{\text{KW}}=a^4\sum_x\Bigg[&\frac{1}{2a}\sum^4_{\mu=1}\Big[\overline{\psi}(x)(\gamma_\mu-i\gamma_4(1-\delta_{\mu4}))U_\mu(x)\psi(x+a\hat{\mu}) \\
    &-\overline{\psi}(x+a\hat{\mu})(\gamma_\mu+i\gamma_4(1-\delta_{\mu4}))U^\dag_\mu(x)\psi(x)\Big]+\overline{\psi}(x)\Bigg(m_0+\frac{3i\gamma_4}{a}\Bigg)\psi(x)\Bigg]
\end{aligned}
\end{equation}
We retain the lattice spacing $a$ explicitly in the action above and in all the subsequent expressions since it will play a role in construction of lattice $\chi$PT. In the momentum space, the free Dirac operator \cite{capitani2010renormalization}  with gauge fields turned off, i.e. $U=1$, is,
\begin{equation}\label{eq:2}
    \mathcal{D}_{\text{KW}}(p)=\frac{i}{a}\sum_{\mu=1}^{4} \gamma_\mu\sin ap_\mu+\frac{i}{a}\gamma_4 \sum_{k=1}^{3}(1-\cos ap_k)+m_0=D(p)+\frac{i}{a}\gamma_4\sum^{3}_{k=1}(1-\cos ap_k)+m_0
\end{equation}
The action preserves chiral symmetry: $\{\mathcal{D}_{\text{KW}},\gamma_5\}=0$. We obtain the momentum space free KW propagator after inverting the free Dirac operator in equation \eqref{eq:2},
\begin{align}\label{eq:3}
        S(p)=\frac{-ia\displaystyle\sum^4_{\mu=1}\gamma_\mu\sin ap_\mu-2ia\gamma_4\sum^3_{k=1}\sin ^2ap_k/2+a^2m_0}{\displaystyle\sum^4_{\mu=1}\sin ^2ap_\mu-4\sin ap_4\sum^3_{k=1}\sin ^2ap_k/2+4\Big(\sum^3_{k=1}\sin ^2ap_k/2\Big)\Big(\sum^3_{l=1}\sin ^2ap_l/2\Big)+(am_0)^2}
    \end{align}
The poles of the free massless propagator are at $(0,0,0,0)$ and $(0,0,0,\pi)$.  Expanding the propagator around the pole $(0,0,0,\pi)$ shows the presence of two degenerate fermionic species having opposite chiralities. The discrete symmetries of KW action are $P$, site-reflection \cite{bedaque2008broken} and $CT$ and so $CPT$.

\section{Lattice Chiral Perturbation Theory}
\label{sec:3}
QCD action in the continuum is the starting point of continuum $\chi$PT \cite{gasser1984chiral}, which is also employed in understanding lattice QCD results. In this work, since we are trying to develop lattice $\chi$PT for MDF, particularly KW fermions on lattice, we begin from the construction of Symanzik action. References \cite{symanzik1983continuum,symanzik1983continuum2} suggest that the lattice action can be written as an expansion in lattice spacing,
\begin{equation}
\begin{aligned}
    S_{\text{sym}}=S_0+aS_1+a^2S_2+... \\
    \text{where, \hspace{0.6cm}} S_{\text{k}}=\sum_i\int d^4xc_i^{(k+4)}O^{(k+4)}_i
\end{aligned}
\end{equation} 
$O^{(k+4)}_i$ are the $(k+4)$-dimensional operators that obey all the symmetries of the lattice action. The operators of all dimensions obeying these symmetries are needed in this construction. \\
 The QCD vacuum breaks the axial flavor symmetry group leading to the spontaneous symmetry breaking of the group  $SU(N)_{\text{A}}$. $N^2-1$ massless Nambu-Goldstone bosons arise from the spontaneous breaking of axial symmetry. These are the lowest-lying pseudoscalar mesons. The quark mass term in the continuum QCD action breaks $SU(N)_{\text{A}}$ explicitly, giving small masses to the pseudoscalar mesons. The low-energy dynamics of QCD are defined by $\chi$PT as the interactions of these pseudoscalar mesons. They are represented as the composite pion field $\Sigma$, defined as $\Sigma=\displaystyle\exp{\big(i\frac{\lambda_{a}\phi_a}{f_0}\big)}$ where $\phi_a$ are the Nambu-Goldstone fields and $\lambda_a$ are the corresponding broken generators. For an introduction to $\chi$PT, we refer the reader to \cite{scherer2003introduction} and references within. 
\section{Single Taste Approach}
\label{sec:4}
The KW propagator has two poles. This means that the fermion wavefunction in the KW action can be broken down to two independent fermion wavefunctions or \textit{tastes}. As a first step, we consider only one taste and construct a lattice $\chi$PT for it.

\subsection{Symanzik Effective Theory}

We select all operators up to dimension-5 that are invariant under the discrete symmetries of the KW action (site-reflection, $P$ and $CT$) and construct the single-taste Symanzik Effective Theory (SET)  up to dim-5.
\begin{align}\label{5}
        S_{\text{sym}}=\int\bigg[a^{-1}\bigg\{&c^{(1)}i\overline{\psi}\gamma_4\psi\bigg\}+a^0\bigg\{\overline{\psi}\gamma_\mu\partial_\mu\psi+\overline{\psi}m\psi\bigg\}\nonumber\\
        +a^1\bigg\{&d^{(0)}\overline{\psi}m^2\psi+d^{(1)}i\overline{\psi}m^2\gamma_4\psi+d^{(2)}\overline{\psi}m\gamma_4\partial_4\psi+d^{(3)}\overline{\psi}m\gamma_j\partial_j\psi\\
        +&d^{(4)}i\overline{\psi}\gamma_4\partial_4\partial_4\psi+d^{(5)}i\overline{\psi}\gamma_4\partial_k\partial_k\psi+d^{(6)}i\overline{\psi}\gamma_l\partial_l\partial_4\psi\bigg\}\bigg]\nonumber
\end{align}
where, summation over repeated Roman alphabets is understood to be the summation over the space directions. The coefficients $c^{(i)}$ and $d^{(i)}$ are independent low-energy constants (LECs).

Spurion analysis is performed by attributing transformation rules to these LECs that make their corresponding term invariant under all the relevant symmetries. These LECs, $c^{(i)}$ and $d^{(i)}$, with their transformation rules and with the lattice spacing dependences are called spurions. As an example, $d^{(1)}$ and quark bilinear $i\overline{\psi}\gamma_4\psi$ transform under $C$, $T$ and $SO(4)$ as,
    \begin{align}
        &d^{(1)}\xrightarrow{C,T}-d^{(1)}\hspace{3cm}i\overline{\psi}\gamma_4\psi\xrightarrow{C,T}-i\overline{\psi}\gamma_4\psi \\
        &d^{(1)}\xrightarrow{SO(4)}\Lambda^{-1}_{\mu 4}d^{(1)}_4\hspace{2.3cm}i\overline{\psi}\gamma_4\psi\xrightarrow{SO(4)}\Lambda_{\mu4}i\overline{\psi}\gamma_4\psi \nonumber
    \end{align}
where $d^{(1)}$ is treated as a four vector with only non-zero component being the fourth component. Using $\Sigma$, its derivatives and together with the spurions, the KW chiral Lagrangian for single fermion species is constructed. The corresponding terms in the Lagrangian at leading orders in $p$, $M$ and $a$ are,
    \begin{align}
        \mathcal{L}_{p^2}&=A_1\langle\partial_k\Sigma\partial_k\Sigma^\dag\rangle+A_2\langle\partial_4\Sigma\partial_4\Sigma^\dag\rangle+A_3\langle M\Sigma+M\Sigma^\dag\rangle \nonumber\\
        \mathcal{L}_{p^2/a^2}&=\frac{1}{a^2}B_1\langle\partial_4\Sigma\partial_4\Sigma^\dag\rangle \label{eq:7}\\
        \mathcal{L}_{p^2a}&=a\big\{\ C_1\langle\partial_k\Sigma\partial_k\Sigma^\dag\rangle+C_2\langle\partial_4\Sigma\partial_4\Sigma^\dag\rangle\big\} \nonumber\\
        \mathcal{L}_{p^2a^2}&={a^2}D_1\langle\partial_4\Sigma\partial_4\Sigma^\dag\rangle \nonumber
    \end{align}
where, the angular brackets stand for trace and the coefficients $A$, $B$, $C$ and $D$ are low-energy couplings not known in principle.

This single fermion model helps us to see how $SO(4)$ symmetry is breaking. It also indicates how to build up the two-taste lattice $\chi$PT without the usual doublers that appear in lattice with fermions. This model is obviously simplistic and far from accurate and, hence, cannot be used for understanding the results from lattice simulations with KW fermions. An additional issue is that the SET we used contains no gluonic terms, which means that it is a free theory. The symmetries of the free and the interacting KW action are crucially different and thus studying the interacting theory is an important step in achieving complete results\footnote{The reason is that in the free KW action, two $U(1)_{\text{A}}$ symmetries are conserved whereas in the interacting KW action, only one survives. We have used the free SET as it is a simpler model to understand.}. 

\section{Two-Taste $\chi$PT for KW fermions}
\label{sec:5}

The next obvious step is to address both the tastes of KW fermions for the construction of the chiral Lagrangian. We separate the fermion wavefunction in equation \eqref{1} into two independent wavefunctions corresponding to the two poles of the KW propagator. This is known as point-splitting \cite{tiburzi2010chiral,creutz2010minimal}. The momentum space KW action coming from the Dirac operator in equation \eqref{eq:2} itself is broken down into two actions corresponding to the two tastes. Point splitting has been discussed in detail by Tiburzi \cite{tiburzi2010chiral} and an alternate version has been provided by Creutz \cite{creutz2010minimal}. For the sake of completeness, we provide a brief summary of Tiburzi's method below. We construct the SET for the free point-split action following which we construct the chiral Lagrangian\footnote{A caveat- we acknowledge that in our work we missed an important ingredient as was pointed out by Steve Sharpe and Stefan Dürr at the Lattice 2024 conference where this work was presented. The principal point is that the total number of pions (Nambu-Goldstone modes) should be three instead of two as has been argued here.}. \\

\subsection{Point Splitting}
\label{subsec:point_splitting}

Here, we recapitulate Tiburzi's approach to point splitting. Once the existence of two fermion tastes is considered, the discrete symmetry transformations ($P$, $C$, $T$, site-reflection) do not preserve the locality of the coordinate space single fermion wavefunction. This is because the momentum support of the wavefunction comes from two different poles with opposite chirality. It is convenient to use a non-local object whose structure of non-locality remains preserved under discrete transformations. \\
First, we divide the Brillouin zone between the tastes by allocating half Brillouin zones to each taste. This is done by defining the two momentum space wavefunctions corresponding to the two tastes as follows,

    \begin{align}
        \psi(k)\big|_{k_\mu\in\mathcal{B}}&=\psi^{(1)}(k)  \\
        \psi(k)\big|_{k_\mu\notin\mathcal{B}}&=\gamma_4\gamma_5\psi^{(2)}(T_{\pi 4}k)\nonumber
    \end{align}
where the new momentum is given by $T_{\pi 4}k_\mu=(\mathbf{k}, (k_4+\pi)\mod{2\pi})$. $\mathcal{B}$ is the central part of the Brillouin zone, $\mathcal{B}:|k_4|<\frac{\pi}{2}$. Two equal parts of the Brillouin zone correspond to the two tastes. We define the isospinor,
\begin{equation}\label{eq:9}
    \Psi(k)=\begin{pmatrix}
                \psi^{(1)}(k) \\
                \psi^{(2)}(k)
            \end{pmatrix}
\end{equation}
In terms of this isospinor, the free two-taste KW action  is,
\begin{equation}\label{eq:10}
    S=\int_{\mathcal{B}}\frac{d^4k}{(2\pi)^4}\overline{\Psi}(k)\bigg[\displaystyle\sum_\mu i(\gamma_\mu\otimes 1)\sin k_\mu-i(\gamma_4\otimes\tau_3)\sum_j(\cos k_j-1)\bigg]\Psi(k)
\end{equation}
The action splits into two parts, each consisting purely of a single taste fermion differing only by the sign of the second term, signifying that the chirality of the two tastes is opposite. The symmetries of the free two-taste action are covered in detail by Tiburzi and are listed as transformations in Tables \ref{tab:1} and \ref{tab:2}. The single taste coordinate space wavefunction loses its locality under the transformations given in Tables \ref{tab:1} and \ref{tab:2}. But the time-smeared wavefunctions $\delta_{\mathcal{B}}\psi_x$ and $\delta_{\mathcal{\overline{B}}}\psi_x$ preserve the structure of their time-smearing under these transformations. The time-smeared wavefunction is defined as,
\begin{equation}
    \delta_{\mathcal{B}}\psi_x=\int_{\mathcal{B}}\frac{dk_4}{2\pi}\displaystyle\sum_{y_4}e^{ik_4(x_4-y_4)}\psi_{\mathbf{x},y_4}
\end{equation}
and similarly for $\delta_{\mathcal{\overline{B}}}\psi_x$ with the integration domain changing to $\overline{\mathcal{B}}$. We say that the quark wavefunction splits into two taste wavefunctions $\delta_{\mathcal{B}}\psi_x$ and $\delta_{\mathcal{\Bar{B}}}\psi_x$.
\begin{table}
\centering
\caption{Transformation properties of time-smeared wavefunctions under discrete symmetries.}
\begin{tabular}{|c|c|c|c|c|c|c|c|c|}
    \hline
     & \textbf{Site-Reflection} & \textbf{Parity} & \textbf{C$\times$T} & \textbf{T$\times$Flavor Rotation} \\
    \hline
    $\delta_{\mathcal{B}}\psi_x$ & $\gamma_4\gamma_5\delta_{\mathcal{B}}\overline{\psi}^T_{\mathbf{1-x},x_4}$ & $\gamma_4\delta_{\mathcal{B}}\psi_{\mathbf{-x},x_4}$ & $\gamma_2\gamma_5\delta_{\mathcal{B}}\overline{\psi}^T_{\mathbf{x},-x_4}$ & $ie^{i\pi x_4}e^{i\theta}\delta_{\mathcal{\overline{B}}}\overline{\psi}_{\mathbf{x},-x_4}$ \\
    \hline
    $\delta_{\mathcal{\overline{B}}}\psi_x$ & $\gamma_4\gamma_5\delta_{\mathcal{\overline{B}}}\overline{\psi}^T_{\mathbf{1-x},x_4}$ & $-\gamma_4\delta_{\mathcal{\overline{B}}}\psi_{\mathbf{-x},x_4}$ & $-\gamma_2\gamma_5\delta_{\mathcal{\overline{B}}}\overline{\psi}^T_{\mathbf{x},-x_4}$ & $-ie^{i\pi x_4}e^{-i\theta}\delta_{\mathcal{B}}\overline{\psi}_{\mathbf{x},-x_4}$ \\
    \hline
\end{tabular}
\label{tab:1}
\end{table}


\begin{table}
\centering
\caption{Transformation properties of the time-smeared wavefunctions under the four $U(1)
$ symmetries.}
\begin{tabular}{|c|c|c|c|c|c|c|c|c|}
    \hline
     & \textbf{Isosinglet Vector} & \textbf{Isovector Axial} & \textbf{Isovector Vector} & \textbf{Isosinglet Axial} \\
    \hline
    $\delta_{\mathcal{B}}\psi_x$ & $e^{i\theta}\delta_{\mathcal{B}}\psi_{x}$ & $e^{i\theta\gamma_5}\delta_{\mathcal{B}}\psi_{x}$ & $e^{i\theta}\delta_{\mathcal{B}}\psi_{x}$ & $e^{i\theta\gamma_5}\delta_{\mathcal{B}}\psi_{x}$ \\
    \hline
    $\delta_{\mathcal{\overline{B}}}\psi_x$ & $e^{i\theta}\delta_{\mathcal{\overline{B}}}\psi_{x}$ & $e^{i\theta\gamma_5}\delta_{\mathcal{\overline{B}}}\psi_{x}$ & $e^{-i\theta}\delta_{\mathcal{\overline{B}}}\psi_{x}$ & $e^{-i\theta\gamma_5}\delta_{\mathcal{\overline{B}}}\psi_{x}$ \\
    \hline
\end{tabular}
\label{tab:2}
\end{table}


\subsection{Two-taste Chiral Lagrangian}
\label{subsec:5.2}

Using the taste wavefunctions of equation \eqref{eq:9} as our fundamental building blocks, the two-taste SET is constructed as given below in equation \eqref{eq:12}. The most general terms made up of the taste wavefunctions are those that are invariant under all the transformations given in Tables \ref{tab:1} and \ref{tab:2}. The action corresponding to the taste originating from the pole at $\{0,0,0,0\}$ is,
\begin{align}\label{eq:12}
        S^{\mathcal{B}}_{\text{Sym}}=\int d^4x \bigg[a^{-1}\big\{&c^{(1)}i\delta_\mathcal{B}\overline{\psi}\gamma_4\delta_\mathcal{B}\psi\big\}\nonumber+a^0\big\{\delta_\mathcal{B}\overline{\psi}\gamma_\mu\partial_\mu\delta_\mathcal{B}\psi+\delta_\mathcal{B}\overline{\psi}m\delta_\mathcal{B}\psi\big\}\nonumber\\
        +a^1\big\{&d^{(0)}\delta_\mathcal{B}\overline{\psi}m^2\delta_\mathcal{B}\psi_x+d^{(1)}i\delta_\mathcal{B}\overline{\psi}_xm^2\gamma_4\delta_\mathcal{B}\psi_x \\
        +&d^{(3)}\delta_\mathcal{B}\overline{\psi}_x m \gamma_4\partial_4\delta_\mathcal{B}\psi_x+d^{(4)}i\delta_\mathcal{B}\overline{\psi}_x \gamma_4 \partial_4 \partial_4 \delta_\mathcal{B}\psi_x\bigg]\nonumber
\end{align}

A similar action $S^{\overline{\mathcal{B}}}_{\text{sym}}$ corresponding to the pole at $\{0,0,0,\pi\}$ can be written by replacing $\delta_{\mathcal{B}}\psi$ with $\delta_{\overline{\mathcal{B}}}\psi$. Hence we see that the SET splits into two SETs each made only of a single taste. Because of the invariance of the total action (sum of the individual single-taste actions) under $T\times$Flavor Rotation symmetry, the coefficients $c^{(i)}$ and $d^{(i)}$ must be equal in both $S^{\mathcal{B}}_{\text{sym}}$ and $S^{\Bar{\mathcal{B}}}_{\text{sym}}$.

The chiral group of the action in equation \eqref{eq:10} comprises of the four $U(1)$ transformations as given in Table \ref{tab:2}. They can be written as,
\begin{equation}
    G=U(1)_V^{\mathcal{B}}\times U(1)_V^{\mathcal{\overline{B}}}\times U(1)_A^{\mathcal{B}}\times U(1)_A^{\mathcal{\overline{B}}}
\end{equation}
This chiral group is spontaneously broken to the group $H=U(1)_V^{\mathcal{B}}\times U(1)_V^{\mathcal{\overline{B}}}$, two axial generators getting spontaneously broken\footnote{This statement is not completely correct. All three generators of $SU(3)_{\text{A}}$ are spontaneously broken while two of them are also explicitly broken. So only one generator of $SU(3)_{\text{A}}$ is preserved explicitly. This assumption leads to the erroneous result of having two pions instead of three, as pointed out by Sharpe and Dürr.}. Therefore, Goldstone's theorem implies the existence of two massless Goldstone bosons. The composite pion field is $\Pi(x)=\exp{(i\phi /f_0})$. Since the broken group is $G/H\equiv U(1)_A^{\mathcal{B}}\oplus U(1)_A^{\mathcal{\overline{B}}}$,
\begin{equation}
    \phi(x)=\pi^a(x)T^a=\pi(x)
    \begin{pmatrix}
            1 & 0 \\
            0 & 0
    \end{pmatrix}
    +\Tilde{\pi}(x)
    \begin{pmatrix}
        0 & 0 \\
        0 & 1
    \end{pmatrix}
    =\begin{pmatrix}
        \pi(x) & 0 \\
        0 & \Tilde{\pi}(x)
    \end{pmatrix}
\end{equation}
where $\pi(x)$ is the pion coming from taste $\mathcal{B}$ and $\Tilde{\pi}(x)$ is the one coming from taste $\overline{\mathcal{B}}$. Thus the composite pion field $\Pi(x)$ becomes,
\begin{equation}
    \Pi(x)=\exp\bigg(i\frac{\phi(x)}{f_0}\bigg)=
    \begin{pmatrix}
        e^{i\pi(x)} & 0 \\
        0 & e^{i\Tilde{\pi}(x)}
    \end{pmatrix}
\end{equation}

\begin{table}[]
    \centering
    \caption{Transformations of the composite pion field.}
    \begin{tabular}{|c|c|c|c|c|c|}
        \hline
         & $\mathbf{P}$ & $\mathbf{C}$ & $\mathbf{T}$ & $\mathbf{G}$ & $\mathbf{T\times}$\textbf{Taste Rotation} \\
        \hline
         $\mathbf{\Pi(x)}$ & $\Pi^\dag(x)$ & $\Pi^T(x)$ & $\Pi(x)$ & $g_{\text{L}}\Pi(x)g_{\text{R}}^\dag$ & $\tau_1\Pi(x)\tau_1$ \\
         \hline
    \end{tabular}
    \label{tab:3}
\end{table}
The transformations of the field $\Pi(x)$ are given in Table \ref{tab:3} where, $g_{\text{L/R}}$ stands for,
\begin{equation}
    g_{\text{L/R}}=\begin{pmatrix}e^{i\theta_{L/R}^{\mathcal{B}}} & 0\\ 0 & e^{i\theta_{L/R}^{{\mathcal{\overline{B}}}}}\end{pmatrix}
\end{equation}
Effectively, $T\times$Taste Rotation exchanges the pion fields that is $\pi(x)\rightleftarrows \Tilde{\pi}(x)$.\\
At various orders in $p$, $a$ and $M$, the terms in the chiral Lagrangian are,
\begin{align}
        \mathcal{L}_{p^2}&=A_1\langle\partial_k\Pi\partial_k\Pi^\dag\rangle+A_2\langle\partial_4\Pi\partial_4\Pi^\dag\rangle+A_3\langle M\Pi+M\Pi^\dag\rangle \nonumber\\
        \mathcal{L}_{p^2/a^2}&=\frac{1}{a^2}B_1\langle\partial_4\Pi\partial_4\Pi^\dag\rangle \label{eq:17}\\
        \mathcal{L}_{p^2a}&=a\big\{\ C_1\langle\partial_k\Pi\partial_k\Pi^\dag\rangle+C_2\langle\partial_4\Pi\partial_4\Pi^\dag\rangle\big\} \nonumber\\
        \mathcal{L}_{p^2a^2}&={a^2}D_1\langle\partial_4\Pi\partial_4\Pi^\dag\rangle \nonumber
\end{align}

The terms appearing in the two-taste chiral Lagrangian are identical to those in equation \eqref{eq:7} except that the structure of the composite pion field is different in each case. This difference brings changes in structure of the interactions of the pions.

\section{Summary}

We have developed a chiral Lagrangian for free KW fermions using two different approaches. The first approach considers only one taste and provides a simplistic model for the eventual development of the full chiral Lagrangian with two tastes. In this approach, symmetries of the single-taste KW action are used to construct a single-taste Symanzik Effective Theory. Spurion analysis is then used to write down the chiral Lagrangian. In the second approach, existence of two tastes is considered. The momentum-space KW action is split into two once the wavefunction is point-split. The symmetries of this two-taste KW action are used to construct a two-taste Symanzik Effective Theory. The pion field is defined following the group structure of the (spontaneously broken) symmetries of the action. Spurion analysis is then performed to obtain the two-taste chiral Lagrangian. The two-taste chiral Lagrangian turns out to be similar to the single-taste chiral Lagrangian except for the structure of the composite pion field.

This work has been done using the free KW action. Complete results will be established using the interacting KW action as the symmetries of the two actions differ significantly. The group arguments used to define the pion field for the two-taste chiral Lagrangian are erroneous and lead to two pions instead of three. For KW fermions, due to lattice discretization errors, one expects the existence of one massless pion and two massive pions in the chiral limit. In the continuum limit, all three pions become massless. Ongoing work focuses on studying the interacting two-taste Symanzik Effective Theory and developing the corresponding chiral Lagrangian. Such a chiral Lagrangian can be used to calculate the masses of the two pions that are massive in the chiral limit as a function of the lattice spacing.
\section*{Acknowledgments}
The authors acknowledge insightful discussions with Prof. Sharpe and Prof. Dürr at the Lattice 2024 symposium. The point regarding the existence of three pions instead of two has been addressed and will appear soon in a paper currently under preparation. In this paper, we have stayed true to what we presented at the Lattice 2024 symposium.

\bibliographystyle{unsrt}
\bibliography{references}

\begin{thebibliography}{10}

\bibitem{gasser1984chiral}
J{\"u}rg Gasser and Heinrich Leutwyler.
\newblock Chiral perturbation theory to one loop.
\newblock {\em Annals of Physics}, 158(1):142--210, 1984.

\bibitem{symanzik1983continuum}
Kurt Symanzik.
\newblock Continuum limit and improved action in lattice theories:(i). principles and $\varphi$4 theory.
\newblock {\em Nuclear Physics B}, 226(1):187--204, 1983.

\bibitem{symanzik1983continuum2}
Kurt Symanzik.
\newblock Continuum limit and improved action in lattice theories:(ii). o (n) non-linear sigma model in perturbation theory.
\newblock {\em Nuclear Physics B}, 226(1):205--227, 1983.

\bibitem{bar2004chiral}
Oliver B{\"a}r, Gautam Rupak, and Noam Shoresh.
\newblock Chiral perturbation theory at o (a 2) for lattice qcd.
\newblock {\em Physical Review D}, 70(3):034508, 2004.

\bibitem{rupak2002chiral}
Gautam Rupak and Noam Shoresh.
\newblock Chiral perturbation theory for the wilson lattice action.
\newblock {\em Physical Review D}, 66(5):054503, 2002.

\bibitem{sharpe1998spontaneous}
Stephen Sharpe, Robert Singleton, et~al.
\newblock Spontaneous flavor and parity breaking with wilson fermions.
\newblock {\em Physical Review D}, 58(7):074501, 1998.

\bibitem{lee1999partial}
Weonjong Lee and Stephen~R Sharpe.
\newblock Partial flavor symmetry restoration for chiral staggered fermions.
\newblock {\em Physical Review D}, 60(11):114503, 1999.

\bibitem{karsten1981lattice}
Luuk~H Karsten.
\newblock Lattice fermions in euclidean space-time.
\newblock {\em Physics Letters B}, 104(4):315--319, 1981.

\bibitem{wilczek1987lattice}
Frank Wilczek.
\newblock Lattice fermions.
\newblock {\em Physical review letters}, 59(21):2397, 1987.

\bibitem{nielsen1981no}
Holger~Bech Nielsen and Masao Ninomiya.
\newblock No-go theorum for regularizing chiral fermions.
\newblock Technical report, Science Research Council, 1981.

\bibitem{creutz2010classification}
Michael Creutz and Tatsuhiro Misumi.
\newblock Classification of minimally doubled fermions.
\newblock {\em Physical Review D—Particles, Fields, Gravitation, and Cosmology}, 82(7):074502, 2010.

\bibitem{boricci2008minimally}
Artan Bori{\c{c}}i.
\newblock Minimally doubled fermion revival.
\newblock {\em arXiv preprint arXiv:0812.0092}, 2008.

\bibitem{borici2008creutz}
Artan Borici.
\newblock Creutz fermions on an orthogonal lattice.
\newblock {\em Physical Review D—Particles, Fields, Gravitation, and Cosmology}, 78(7):074504, 2008.

\bibitem{creutz2008four}
Michael Creutz.
\newblock Four-dimensional graphene and chiral fermions.
\newblock {\em Journal of High Energy Physics}, 2008(04):017, 2008.

\bibitem{creutz2008local}
Michael Creutz.
\newblock Local chiral fermions.
\newblock {\em arXiv preprint arXiv:0808.0014}, 2008.

\bibitem{capitani2010renormalization}
Stefano Capitani, Michael Creutz, Johannes Weber, and Hartmut Wittig.
\newblock Renormalization of minimally doubled fermions.
\newblock {\em Journal of High Energy Physics}, 2010(9):1--26, 2010.

\bibitem{bedaque2008broken}
Paulo~F Bedaque, Michael~I Buchoff, Brian~C Tiburzi, and Andre Walker-Loud.
\newblock Broken symmetries from minimally doubled fermions.
\newblock {\em Physics Letters B}, 662(5):449--455, 2008.

\bibitem{scherer2003introduction}
Stefan Scherer.
\newblock Introduction to chiral perturbation theory.
\newblock In {\em Advances in Nuclear Physics, Volume 27}, pages 277--538. Springer, 2003.

\bibitem{tiburzi2010chiral}
Brian~C Tiburzi.
\newblock Chiral lattice fermions, minimal doubling, and the axial anomaly.
\newblock {\em Physical Review D—Particles, Fields, Gravitation, and Cosmology}, 82(3):034511, 2010.

\bibitem{creutz2010minimal}
Michael Creutz.
\newblock Minimal doubling and point splitting.
\newblock {\em arXiv preprint arXiv:1009.3154}, 2010.

\end{thebibliography}

\end{document}